\documentclass[journal=nalefd,manuscript=article]{achemso}

\usepackage[version=3]{mhchem} % Formula subscripts using \ce{}
\usepackage{natbib}

\title{Molecular beam growth of graphene on mica}

\author{Gunther Lippert}
\email{lippert@ihp-microelectronics.com}
\phone{+49 (0)335 5625727}
\fax{+49 (0)335 5625681}
\affiliation[IHP]
{IHP, Im Technologiepark 25, 15236 Frankfurt (Oder), Germany}

\author{Jarek D\k{a}browski}
\author{Yuji Yamamoto}
\affiliation[IHP]
{IHP, Im Technologiepark 25, 15236 Frankfurt (Oder), Germany}

\author{Felix Herziger}%

\author{Janina Maultzsch}% 

\affiliation [Technische Universit\"at Berlin] {Institut f\"ur Festk\"orperphysik, Technische Universit\"at Berlin, Hardenbergstr. 36, 10623 Berlin, Germany}

\author{Max C. Lemme}%
\affiliation [KTH] {KTH Royal Institute of Technology, Isafjordsgatan 22, 16440 Kista, Sweden}
\author{Wolfgang Mehr}
\author{Grzegorz Lupina}
\affiliation[IHP]
{IHP, Im Technologiepark 25, 15236 Frankfurt (Oder), Germany}

\begin{document}
\begin{abstract}
We demonstrate molecular beam growth of graphene on biotite mica substrates at temperatures below $1000^\circ$C. As indicated by optical and atomic force microscopy, evaporation of carbon from a high purity solid-state source onto biotite surface results in the formation of single-, bi-, and multilayer graphene with size in the micrometer regime. It is shown that the graphene grown directly on mica surface is of very high crystalline quality with the defect density below the threshold detectable by Raman spectroscopy. The interaction between graphene and the mica substrate is studied by comparison of the Raman spectroscopy and atomic force microscopy data with the corresponding results obtained for graphene flakes mechanically exfoliated onto biotite substrates. Experimental insights are combined with density functional theory calculations to propose a model for the initial stage of the van der Waals growth of graphene on mica surfaces. This work provides important hints on how the direct growth of high quality graphene on insulators can be realized in general without exceeding the thermal budget limitations of Si technologies.

\end{abstract}
%\textbf{KEYWORDS}: Graphene, growth on insulators, van der Waals, growth models, Raman spectroscopy, density functional theory

\vspace{0.2in}

One of the biggest obstacles on the way to various technological applications of graphene is still the lack of a fabrication method for %large area 
high quality graphene on arbitrary substrates \cite{lundstrom2011, editorial}. Although remarkable progress has been recently made in catalytic growth and transfer of large graphene layers on Cu \cite{Yu, Li, liang2011}, some important problems remain unsolved. For instance, the transfer process from Cu onto desired substrates often results in the contamination of the graphene surface or the formation of folds and ripples, which degrade its mechanical and electronic characteristics \cite{Schultz}. As a result, other manufacturing methods enabling direct deposition of high quality graphene are intensively investigated \cite{samsungnature11}. One of the alternatives is the direct molecular beam growth (MBG) using molecular beam epitaxy equipment (MBE). MBG has the potential to provide intrinsically purer graphene as compared to other methods and better thickness control. Furthermore, MBG does not involve catalytic processes and thus can be performed on various substrates, including semiconductors and insulators which are of high technological relevance for many applications. In the literature, there are only few reports on attempts to grow graphene by MBG on a relatively large variety of substrates including Si \cite{hackley,maeda}, SiC \cite{moreau, park}, hexagonal boron nitride \cite{garcia1}, muscovite mica \cite{lippert,garcia1,jerng2011,wurstbauer2012}, \ce{SrTiO3} \cite{jerng2011}, YSZ \cite{jerng2011}, and sapphire \cite{jerng,maeda2}. A broad range of substrate temperatures ($200-1400^\circ$C) was used and various carbon sources such as pyrolytic graphite filaments \cite{hackley,moreau,park,garcia1,lippert,jerng2011,jerng}, C$_{60}$\cite{park}, and gas sources \cite{maeda,maeda2} were applied. These studies, depending on the chosen substrate, temperature, and carbon source, report the formation of either carbides, graphitic carbon, nanocrystalline graphite, or multilayer graphene. However, the direct molecular beam growth of high quality single layers of graphene on technologically relevant substrates and at moderate temperatures ($<1000^\circ$C) remains a great challenge.
 
Mica, a silicate-based mineral, has been often used as a substrate for the growth of various materials \cite{grunbaum, krakow, lamellas}. In particular, mica, with no dangling bonds on its clean surface, is one of the preferred substrates for van der Waals epitaxy which enables formation of heterostructures composed of materials with a large lattice mismatch\cite{koma1}. The van der Waals epitaxy was in the past successfully applied to grow high quality thin films of two-dimensional layered crystals such as \ce{MoS2} and \ce{MoSe2} \cite{ueno} and thin layers of $\alpha$-alumina \cite{steinberg} on mica surfaces. In the context of graphene research, mica is considered as a candidate substrate material for graphene device prototyping due to its atomic smoothness and excellent insulating properties \cite{rudenko2011, lui}.
 
In this Letter, we present a van der Waals epitaxy-based approach enabling direct growth of graphene layers on mica substrates at temperatures below $1000^\circ$C. 
In our previous growth experiments, \cite{lippert} performed mainly on low grade muscovite mica, the formation of multilayer graphene flakes has been observed. The growth of graphene/graphite islands at elevated temperatures ($>850^\circ$C), which favor the growth of flakes with high crystalline quality, was accompanied by local decomposition of the substrate and formation of highly distorted carbon (or quasi-highly reduced graphene oxide\cite{ramesha2009,moon_rgo}).
Here, a high grade biotite mica with superior thermal stability \cite{tlapek} is used as the substrate. Using optical and atomic force microscopy, we show that evaporation of carbon onto biotite surface results in the formation of single-, bi-, and multilayer graphene layers with the size in the micrometer regime. Based on Raman spectroscopy it is demonstrated that the single layer graphene grown directly on mica surface is of very high crystalline quality. Furthermore, we compare the graphene obtained by van der Waals growth with graphene flakes deposited by mechanical exfoliation on the surface of biotite mica. These results constitute a significant progress in the field of direct graphene growth on insulating substrates at moderate temperatures.

Biotite mica (K(Mg,Fe)$_{3}$AlSi$_{3}$O$_{10}$(OH,F)$_{2}$) with monoclinic structure, space group $C2/c$ and cleavage on the ($001$) face is used as the substrate \cite{baba}. The growth of graphene layers is accomplished in a molecular beam epitaxy chamber equipped with a solid state carbon source (Figure 1a). Mica samples are attached to the sample holder (Figure 1b) and heated up to the target temperature ($500-1000^\circ$C). The carbon source (Figure 1c) is composed of high purity pyrolytic graphite filament heated by passing a high current ($\sim 110$A) to temperatures of $\sim 2500^\circ$C. The growth rate is around 1\AA/min. The chemical composition of the sample during graphene growth was monitored in-situ by x-ray photoelectron spectroscopy (XPS). The samples were then analyzed ex-situ by $\mu$-Raman spectroscopy, atomic force microscopy (AFM) and Auger Electron Spectroscopy (AES). 

Figure $1$d shows the overview XPS scan from the biotite sample obtained immediately after cleavage in ambient air. The presence of Al, Si, O, K, Mg, Fe, and F photoemission peaks confirms the typical chemical composition of biotite \cite{nicolini}.
\begin{figure}[h]
\includegraphics[width=0.5\columnwidth]{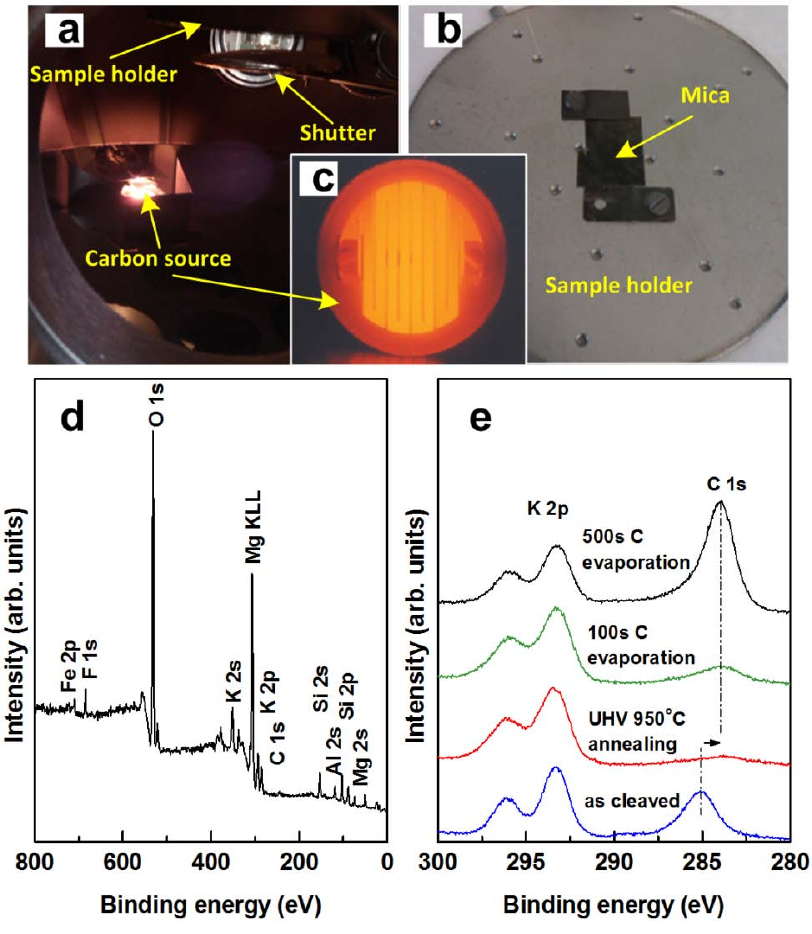}

\vspace{-10pt}

\caption{\label{fig:xps_new} Experimental set-up and x-ray photoemission study of graphene growth on mica. (a) look into the growth chamber through a viewport with visible carbon source and sample holder, (b) mica sample mounted on the sample holder, (c) magnified image of the glowing pyrolytic carbon source (d) overview spectrum of the substrate after cleavage in air, (e) K $2p$ - C $1s$ spectral region at various experimental stages.}
\end{figure}
Figure $1$e shows high resolution photoemission spectra around the K $2p$ - C $1s$ region measured at various experimental stages. The as-cleaved biotite sample shows the C $1s$ peak at a binding energy of about $285$ eV which can be associated with oxidized carbon \cite{xiaolinLi}. After UHV annealing, performed at $950^\circ$C for $10$ min, the C $1s$ peak significantly decreases in intensity which can be attributed to the desorption of loosely bonded C-O species \cite{poppa}. The UHV annealing fails, however, to remove the C $1s$ emission completely. The weak residual signal observed at the C $1s$ position can be due to the presence of   carbonaceous compounds which are tightly bound to the surface and are difficult to remove even with a  prolonged annealing at elevated temperatures \cite{ostendorf}. Evaporation of carbon onto the annealed mica substrate results in a new XPS peak centered at a binding energy of around $284$ eV which is characteristic for C-C bonds \cite{xiaolinLi}. This peak clearly gains intensity with increasing evaporation time giving evidence that the amount of C-C bonds on the surface of mica increases during evaporation.

\begin{figure}[h]
\includegraphics[width=0.5\columnwidth]{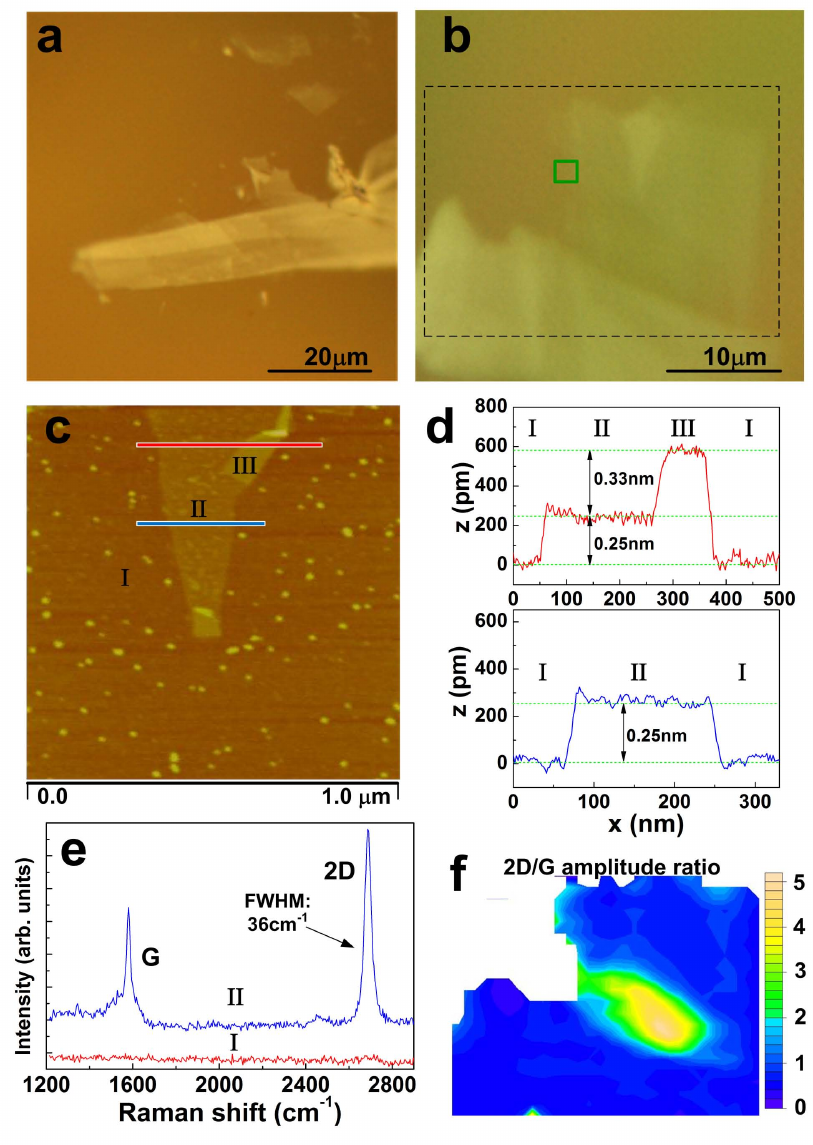}

\vspace{-10pt}

\caption{\label{fig:microscope} Characterization of graphene flakes deposited on mica by carbon evaporation. Optical microscope image of a carbon island with low (a) and high (b) transparency. (c) AFM height-mode image of the region marked by the green rectangle in panel (b). Mica surface (I) and areas covered by single (II) and bi-layer (III) graphene can be observed. (d) AFM step height scans across the areas of mica substrate (I), single layer graphene (II), and bi-layer graphene (III); marked by red and blue lines in panel (c). (e) Raman spectra measured at $514$ nm excitation wavelength from the areas I and II. (f) Raman 2D/G map from the selected area of a highly transparent region as marked by the black rectangle in panel (b).}
\end{figure}

Figure $2$a shows a representative optical microscope image taken after evaporation of carbon for $50$s onto the mica substrate at $950^\circ$C. Randomly distributed features with varying transparency and size, as those shown in Figure $2$a, can be observed on the entire sample area (about $20 \times 20$mm). According to scanning Auger Electron Spectroscopy (AES, which has superior lateral resolution as compared to XPS) the irregular islands on the biotite surface are composed exclusively of carbon and are separated from one another with exposed areas of mica. With increasing deposition time, lateral and vertical dimensions of the islands increase. As we will show below, these islands are monolayer, bilayer and multilayer graphene. However, even at very long deposition times (several minutes) large areas of the substrate do not show any significant coverage with carbon according to AES. This suggests that the incoming carbon atoms are very mobile on the mica surface and tend to agglomerate around certain growth centres. Formation of such agglomerates can be clearly observed only at elevated substrate temperatures (>$800^\circ$C). When the substrate temperature is below this regime, on the other hand, the surface of biotite is rather covered with a relatively uniform layer of amorphous carbon. This is similar to the behavior which we previously observed for muscovite mica \cite{lippert}. For deposition at relatively high substrate temperatures ($900-1000^\circ$C) it can be seen that the islands formed on biotite exhibit a layered structure with a clear optical contrast between various terraces (see Figure $2$a). Besides thicker islands, like the one seen in Figure $2$a, there are many thinner and smaller islands with high transparency. In optical microscope images, the latter features closely resemble the graphene flakes deposited on muscovite substrates by mechanical exfoliation \cite{lui}. Figure $2$b shows a higher magnification optical micrograph of a feature with high transparency. The edge of this island (area marked in green) was investigated by AFM (Figure $2$c-d). The AFM topography image displayed in Figure $2$c shows a triangular feature which extends from the island in Figure $2$b. Figure $2$d presents the results of step height scans along the blue and red lines (marked in Figure $2$c) across the mica substrate (I) and covered areas with high (II) and low transparency (III).
The step height between the mica substrate and area II is about $0.25$ nm. The step height between areas II and III is around $0.33$ nm, which is in a very good agreement with the interlayer distance in graphite \cite{Franklin}. Based on these AFM results, and Raman spectroscopy (see below), we argue that the areas II and III correspond to the single and bi-layer graphene, respectively, which are formed on the mica substrate as a result of carbon evaporation. 
To gain more insight into the interaction of the MBG graphene with the substrate, the AFM measurements in Figure $2$d were compared with those obtained for exfoliated single and bi-layer graphene on biotite mica. While the distance between the first and the second graphene layer is in both cases very similar, the step height between the mica substrate and the first graphene layer is for the MBG sample about $3$ times smaller than for the exfoliated flakes (measured with the same AFM set-up). This indicates that the interaction of the MBG graphene layer with the biotite surface is stronger than in the case of exfoliated layers. The distance of the order of 0.3\,nm is consistent with that obtained 
in ab initio calculations (Ref.\cite{rudenko2011} and this work), if we assume that it reflects the distance between the C atoms and the topmost O atoms of mica and that (the majority of) K atoms are missing. With all K atoms present, the expected distance is 0.49\,nm \cite{rudenko2011}, which is still less than around 0.7\,nm usually observed for exfoliated graphene flakes on mica \cite{xu}. This may indicate that the step between mica and single layer graphene is higher for exfoliated samples due to some "dead" space under the graphene sheet, possibly caused by molecules physisorbed on mica surface exposed to air. The small features visible as bright spots distributed over mica substrate can be observed also on unprocessed surfaces \cite{ostendorf2009}. We thus suppose that this are nano-particles of a potassium compound (oxide, hydroxide, or carbonate).

Figure $2$e displays the micro-Raman spectra obtained with the laser spot focused on the areas I and II. In area II, clear Raman G and $2$D peaks (full width at half maximum (FWHM) of the $2$D mode $\sim 36$ cm$^{-1}$) of graphene can be observed. Due to the limited experimental lateral resolution, signals originating from the single (II) and double (III) graphene layer cannot be clearly resolved and the measured spectrum is most probably the result of overlapping contributions from areas II and III. In contrast, Raman measurements on the mica substrate (I) do not give any Raman signatures in the investigated spectral range. 
In search for larger areas of single-layer graphene, a Raman mapping has been performed on a selected region of the island shown in Figure $2$b (black rectangle). Based on these data, the ratio of the 2D and G peak amplitudes (2D/G) was calculated. The results are shown in Figure $2$f. High 2D/G ratios of up to $5$ and narrow 2D peaks (FWHM $<30$ cm$^{-1}$) indicate the presence of a relatively large ($\sim$ $1$x$5\mu$m) single layer graphene sheet. 
In the following, we focus on a more detailed analysis of the Raman spectra originating from the latter region and compare them to the spectra obtained from graphene deposited on biotite surface using mechanical exfoliation.

Figure 3a shows the Raman spectrum obtained with 532\,nm laser excitation wavelength from the MBG graphene sample shown in Figure $2$. As mentioned above, the well-known characteristic G and 2D peaks indicate that the MBG samples consist of graphene-like \textit{sp}\textsuperscript{2} carbon. The absence of the disorder-induced D mode at $\sim$1350\,cm\textsuperscript{-1} reflects the very high crystalline quality of the layers. The D mode was sometimes observed at edges of the grown graphene samples. 
\begin{figure}[h]
\includegraphics[width=0.4\columnwidth]{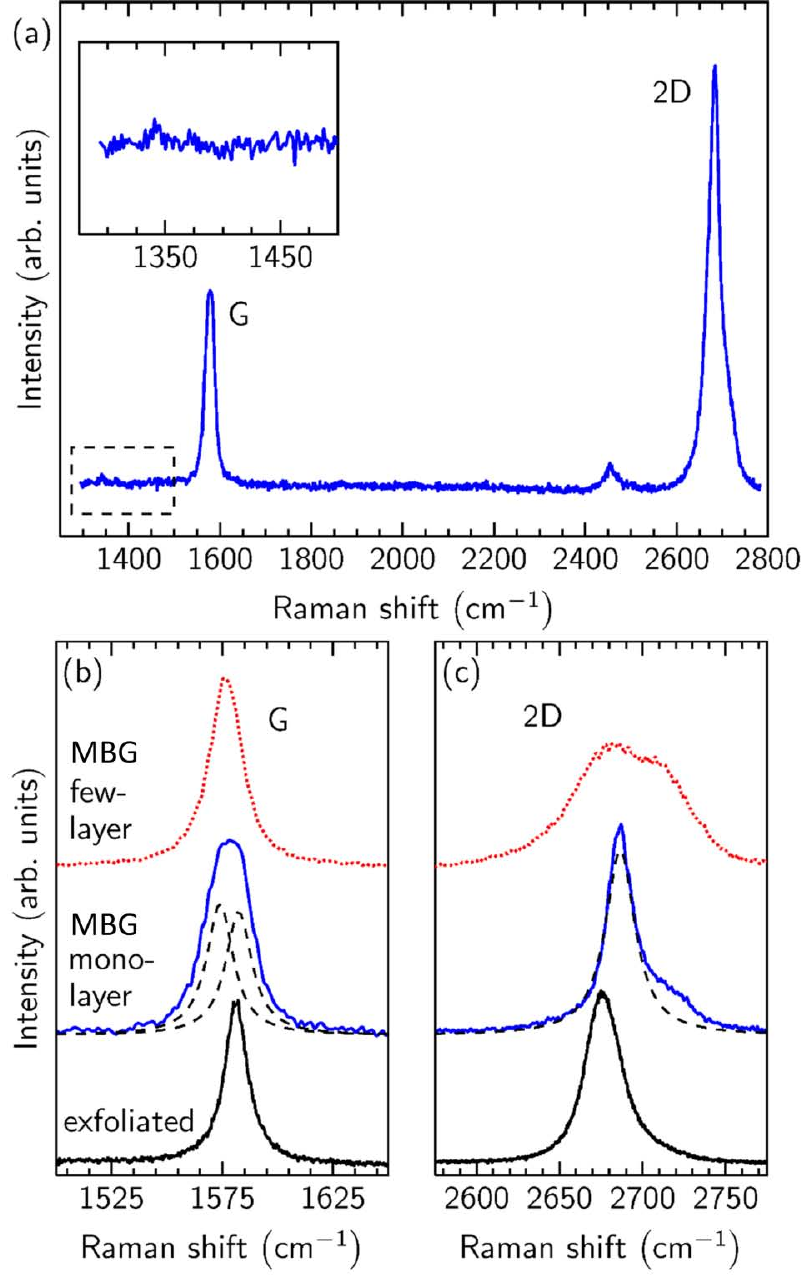}

\vspace{-10pt}

\caption{\label{fig:raman}(a) Raman spectra of MBG graphene on mica acquired with 532 nm laser excitation wavelength. The inset shows the D-peak region, where no defect-induced peak can be observed. A higher magnification of the G and 2D peaks of this spectrum is shown in panels b and c (blue lines), respectively. (b) and (c) G and 2D band, respectively, of the MBG few-layer graphene (upper spectra), MBG monolayer (middle spectra) and exfoliated monolayer graphene (lower spectra) on mica. The dashed lines are Lorentzian fits. The 2D band of exfoliated graphene on mica shows a slightly asymmetric shape. Spectra are vertically offset for clarity.}%
\end{figure}
The G band at 1580\,cm\textsuperscript{-1} (Figure 3b) is composed of two single Lorentzians (at 1574\,cm$^{-1}$ and 1582\,cm$^{-1}$) with FWHM of $\sim$14\,cm\textsuperscript{-1}.
The 2D band of the MBG graphene in Figure 3c consists of a single Lorentzian with FWHM $\sim$22\,cm\textsuperscript{-1} and an additional feature on the high frequency side. This peak shape is distinctly different from bilayer or multilayer 2D band, where at least four contributions are present (two with approximately equal amplitude) with typical FWHM of 25 - 30\,cm$^{-1}$. 
We therefore conclude that the narrow 2D band of the MBG graphene originates from a single graphene layer. The 2D shoulder observed at higher frequencies is most probably due to an overlap of Raman signals from the monolayer and the surrounding few-layer graphene.

The 2D frequency of the MBG graphene on mica ($\sim$2686\,cm$^{-1}$) is slightly increased with respect to exfoliated graphene on mica ($\sim$2676\,cm\textsuperscript{-1}) and on SiO$_2$ ($\sim$2675\,cm\textsuperscript{-1}). The difference in 2D-peak positions between the MBG and exfoliated samples on mica have to be caused by the growth process and is discussed below. 

We now consider the G band of the MBG graphene shown in Figure 3b. As mentioned above, we observe two Lorentzians in the G band (middle curve), in contrast to the single peak observed in exfoliated monolayer graphene (lower curve). This shows that the two peaks in the G band of the MBG graphene are not induced by the substrate. We discuss the following possible origins (\textit{i}) strain, (\textit{ii}) doping, and (\textit{iii}) overlap of single and few-layer graphene.

(\textit{i}) We first consider uniaxial strain as a reason for the G-peak splitting and compare our data with measurements from Mohiuddin \textit{et al.} \cite{mohiuddin2009} and calculations of Mohr \textit{et al.} \cite{mohr2009}. The measured splitting of 8 cm$^{-1}$ would correspond to a strain value of $\sim 0.4$\%. In this case, one should additionally observe a downshift by -5\,cm$^{-1}$ and -13\,cm$^{-1}$, for the low and high frequency components of the G band, respectively. The measured peak shift with respect to the exfoliated graphene G peak is 0\,cm$^{-1}$ and -8\,cm$^{-1}$ and hence below the theoretical predictions. Furthermore, strain should shift the 2D band by about -25\,cm$^{-1}$, which is not observed in our investigations. In contrast, the 2D mode is slightly blue-shifted to $\sim$2686\,cm\textsuperscript{-1} with respect to the 2D mode of exfoliated monolayer graphene on mica (Figure 3c). Therefore, uniaxial strain (\textit{e.g.}, resulting from the growth process) cannot explain the two observed Lorentzians in the G band. 

(\textit{ii}) When graphene is grown on mica substrates, it is
likely that doping occurs due to for example nonuniform distribution of potassium atoms on the substrate surface \cite{rudenko2011}. Doping cannot describe the observed splitting of the G band, but can give an explanation for the reduced FWHM \cite{pisana2007}. The two Lorentzians in Figure 3b have a FWHM of $\sim$14\,cm\textsuperscript{-1}, which is slightly below the usual value of undoped monolayer graphene. Quantitatively, this narrowing is consistent with low doping levels on the order of 10$^{12}$ - 10$^{13}$\,cm$^{-2}$, which in turn would explain the presence and magnitude of the 2D-band shift \cite{Adas2008}. The direction of the shift would in this case mean that graphene is p-doped  \cite{Adas2008}, that is, that the surface is more electronegative than graphene, or some acceptor species are adsorbed on graphene, or both. The former can be due to removal or electrical neutralization of a certain amount of K atoms, the latter may be due to the presence of residual oxygen or OH in the MBE chamber. 

In exfoliated graphene on mica (Figure 3b bottom spectrum), the G mode has slightly smaller width than in the MBG graphene, indicating even higher doping. As a consequence, we would expect that the 2D mode in exfoliated graphene on mica has similar or higher frequency than in MBG graphene. This is, however, not the case. Instead, the 2D-mode frequency in exfoliated graphene on mica is similar to that in exfoliated graphene on \ce{SiO2}. Moreover, the G-mode frequency in our MBG graphene is rather low compared to recently published data on p-doped graphene exfoliated on mica \cite{shim2012}. 
Therefore, we believe that, in addition to p-doping by the substrate, the electron band structure of MBG graphene is affected by the growth process. As due to double resonance the 2D mode is much more sensitive to modifications of the band structure than the G mode \cite{thomsen2000, maultzsch2004}, this would explain why the G mode shows very little change compared to slightly doped graphene, whereas the 2D mode is strongly affected.

(\textit{iii}) The above discussed 2D-band lineshape was explained by two Raman signals overlapping. By comparing the MBG monolayer Raman spectrum of the G band with a Raman spectrum of surrounding few-layer graphene [see Figure 3b, upper curve], we observe that only the low-frequency peak remains visible. Following the idea of two Raman signals overlapping, we assign the higher-frequency G peak in Figure 3b to monolayer graphene and the lower-frequency peak to surrounding few-layer graphene. Indeed, the G band shifts with the number of graphene layers to lower frequencies, although by a smaller amount than observed here. \cite{gupta2006, das2008} As a consequence, we can explain the two Lorentzian peaks that form our G band also by overlap.

The above Raman investigations prove that the studied MBG sample consists of a graphene monolayer which is surrounded by a few-layer graphene. The absence of a D band in the monolayer region demonstrates the high crystalline quality of the deposited layer.

We now return to the observation of phase separation between graphitized areas and clean mica surface. This separation is not obvious to occur, because atomic carbon deposited from the MBE source is expected to react chemically with the surface to form relatively stable carbidic species, which are likely to be immobile at temperatures that mica can withstand. This rises the question about the physical mechanism of carbon transport from reaction sites on mica to graphitic islands. In order to clarify this issue, we performed ab initio density functional theory calculations (for details see SI) for the initial stage of carbon deposition on mica.

\begin{figure}[h]
\includegraphics[width=0.45\columnwidth]{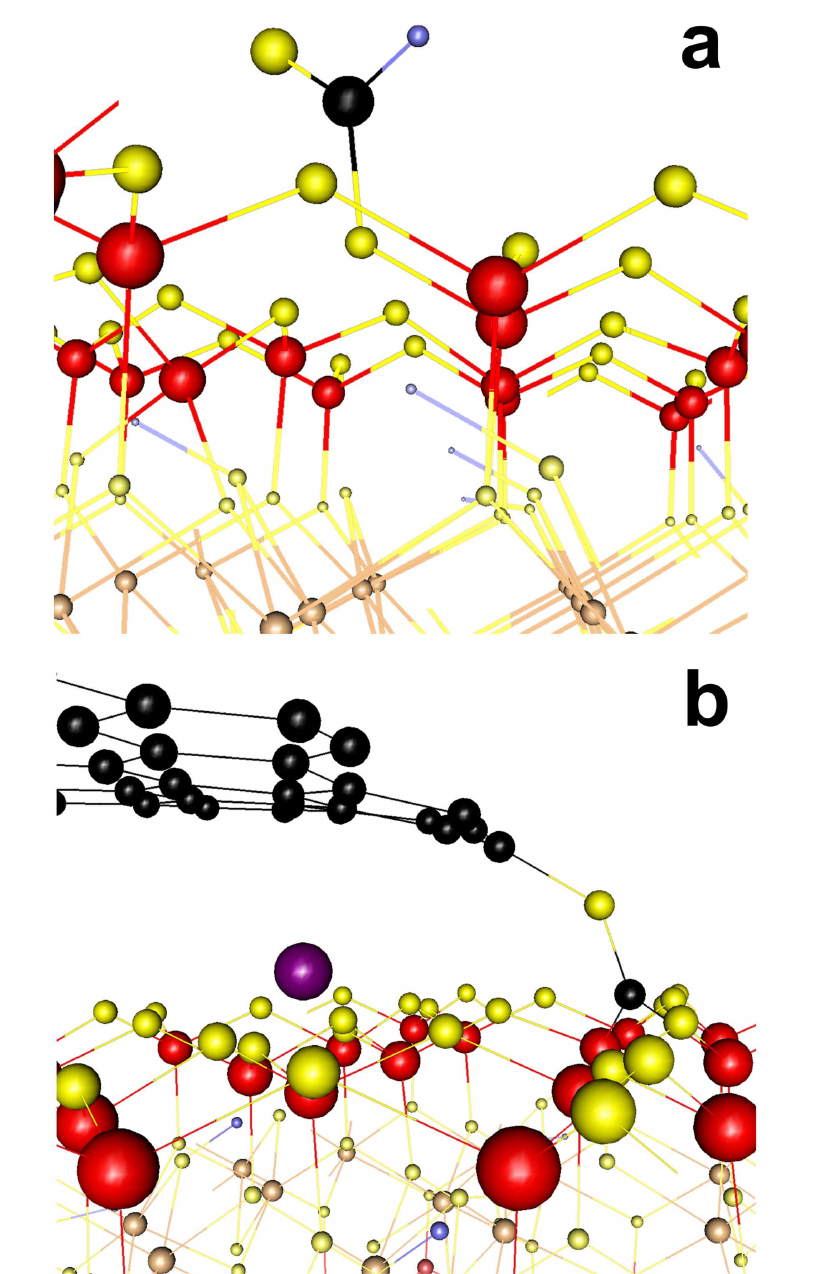}

\vspace{-10pt}

\caption{\label{abinitio} Selected structures pertinent to carbon transport on mica surface. Carbon is black, silicon is red, oxygen is yellow, hydrogen is blue, aluminum is orange, potassium is purple. (a) Carbon in excited formyl species. (b) Oxygen bridge between CO and a graphene molecule.}
\end{figure}

Results of the calculations confirm that atomic carbon reacts with the mica surface. This either reduces the surface producing volatile CO and subsequently substitutional carbon defects, C$_{\rm O}$, or generates chemisorbed formyl (HCO) species, whereby oxygen is not taken away from the surface but remains bonded in formyl moieties. C$_{\rm O}$ defects do not seem to be mobile. The formyl species are potentially mobile via excitation from the ground state to the excited state (Figure 4a). The energy difference between these states amounts to $1.55$ eV, meaning that if no additional energy barrier existed on the way from the ground to the excited state and if the excited state could jump between oxygen atoms without any energy barrier, then during $1$ ms the carbon-carrying formyl could migrate about $10$ nm away from the original reaction site (assuming a random walk process and attempt frequency of $10^{13}$ s$^{-1}$). Even with this optimistic estimate, which ignores atomistic details of the migration process, formyl migration would not lead to efficient transport of carbon from the mica surface to graphitic islands.
On the other hand, ab initio molecular dynamics shows that at growth temperatures small graphene molecules (clusters of cyclic carbon) can migrate on mica surface with thermal speed, i.e. effectively with no diffusion barrier. We verified this for a graphene molecule built of $24$ carbon atoms. The motion is barrierless because graphene is incommensurate with the substrate (this levels out the energy surface) and because temperature excites out-of-plane vibration of graphene (this helps to overcome local hills on the energy surface). These mobile graphene molecules can collect formyl species and also incorporate carbon from C$_{\rm O}$. They also help to recover the stoichiometry of the surface. Namely, the edge of a graphene molecule is likely to contain numerous O atoms and OH species from consumed formyl and from residual O and OH in the reaction chamber. Energy (about $2.2$ eV) is gained when oxygen is returned from a graphene edge to a surface oxygen vacancy produced when carbon is moved from C$_{\rm O}$ to graphene. Moreover, it appears that the presence of oxygen species on the edge of graphene molecules facilitates the process of collection of carbon from C$_{\rm O}$ defects. Oxygen enables a multi-step reaction path in which the first step (barierless formation of oxygen bridge between graphene and C$_{\rm O}$, cf. Figure 4b) provides the energy to overcome the barrier on the way to complete the reaction (exchange the O atom and the C atom from the C$_{\rm O}$ defect). The reaction in which an oxygen atom adsorbed at the graphene edge replaces carbon in C$_{\rm O}$ and the carbon atom becomes bonded to the graphene edge releases about $4.4$ eV.
It is noteworthy that \ce{SiO2} surfaces are also terminated with oxygen bonded to two Si atoms, as is the case for mica. This suggests that the mechanism sketched above can also apply to \ce{SiO2} substrates, albeit the energy landscape may be somewhat different because the bonding between silicon and oxygen on \ce{SiO2} is more complex than on mica.

In summary, we have demonstrated a method based on van der Waals epitaxy enabling direct growth of single, bi-, and multilayer graphene on mica substrates at temperatures not exceeding $1000^\circ$C. Raman spectroscopy was used to identify graphene layers with sizes in the micrometer range and to prove their very high crystalline quality. With successful further optimization of the growth process, this method can open the way to direct local deposition of graphene on insulating substrates in technological applications. While mica will ultimately not be used for the integration with Si microelectronics, it may, well serve as an ideal substrate for prototyping new generations of graphene devices. A recent report on the preparation of atomically thin mica layers on \ce{SiO2}/Si substrates \cite{gomez} and the encouraging results on the synthesis of thin mica films by vapour phase methods \cite{nakasone} underline the potential of this material as an ultrathin insulator and substrate for graphene-based electronic devices.

Furthermore, our results in conjunction with recent reports on successful MBE growth of atomically smooth oxide tunnel barriers \cite{wang} and topological insulators \cite{chen} on graphene show that molecular beam growth is an important method for exploring prototypes of advanced graphene-based devices.

\acknowledgement

Atomistic calculations have been done at the J\"ulich Supercomputing Centre, Germany, NIC project hfo06. The authors thank I. Costina, O. Fursenko, A. Georgi, S. Hasselbach, T. Schroeder, P. Zaumseil, and M. Z\"{o}llner for experimental support and discussions. JM acknowledges support from the European Research Council, ERC grant no. $259286$.

\newpage

\bibliography{references}

\end{document}